\documentclass[12pt]{iopart}  

 \usepackage{amssymb} 
  \usepackage{setstack}  
 \usepackage{geometry}
\usepackage{graphicx}
\usepackage{epstopdf}
\usepackage{bm}

\def\AOM {acousto-optic modulator}

\def\ICR{inverted crossover resonance}
\def\ICL {intercombination line}
\def\LIA {lock-in amplifier}
\def\MOT {magneto-optical trap}

\def\ppKTP {periodically poled potassium titanyl phosphate}

\def\PMT {photo-multiplier tube}

\def\SAS{saturated absorption spectroscopy}

\def\SNR {signal-to-noise ratio}

\def\VCO {voltage controlled oscillator}
\def\wrt {with respect to}

\def\ARC {Australian Research Council} 

\newcommand{\degC}{$^{\circ}$C}

\newcommand{\si}{$\sim$}

\newcommand{\uK}{$\mu$K}

\newcommand{\fastT}{$^{1}S_{0}-\,^{1}P_{1}$}    
\newcommand{\coolingT}{$^{1}S_{0}-\,^{3}P_{1}$}  

\newcommand{\clockT}{$^{1}S_{0}-\,^{3}P_{0}$}

\newcommand{\clockTdash}{$^{1}S_{0} -\,^{3}P_{0}$} 

\newcommand{\Yb}{$^{171}$Yb}	
\newcommand{\Ybtwo}{$^{172}$Yb}

\begin{document}


\title[]{Inverted crossover resonance within one Zeeman manifold} 

\author{ L. A. Salter$^{1}$, E. de Clercq$^{2}$  and  J.~J.~McFerran$^{1}$ } 
\address{$^{1}$School of Physics, University of Western Australia, 6009, Crawley, Australia}
\address{$^{2}$LNE-SYRTE, Observatoire de Paris, PSL Research University, CNRS, Sorbonne Universit\'es, UPMC Univ. Paris VI, 61 avenue de l'Observatoire, 75014 Paris, France}
\ead{john.mcferran@uwa.edu.au}

\begin{abstract}

We carry out investigations of inverted crossover resonances in $\pi$-driven four-level systems where  $\Delta F$ can be zero. Through the use of sub-Doppler frequency modulation spectroscopy  of  the $(6s^{2})$~$^{1}S_{0}$ $-$  $(6s6p)$\,$^{3}P_{1}$  transition in $^{171}$Yb the resonance becomes manifest. The centre-frequency is inherently insensitive to first-order Zeeman shifts and equates to the two-level resonance frequency in the absence of a magnetic field.  A rate equation model is used to help validate the nature of the resonance. Optical frequency measurements of the $F'=1/2$ hyperfine line recorded over two months demonstrate  a statistical uncertainty of $2\times10^{-11}$.  The inverted crossover resonance found with the $F'=3/2$ line is used for 556\,nm laser frequency stabilization,  which is an alternative means when applied to magneto-optical trapping of $^{171}$Yb.



\end{abstract} 



                         
                           
                            


\maketitle 


\vspace{0.5cm}

  
\section{Introduction}
Explorations in degenerate Fermi gases~\cite{Fuk2007a,Han2011,Dor2013},  quantum many body simulations with lattices~\cite{Sca2014, Sch2015a,Yam2016}, 
and atomic clocks~\cite{Por2004a,Hin2013,Bel2014,Tak2015,Nem2016} each make use of neutral ytterbium.  A common approach to producing atomic cloud temperatures at a few tens of microkelvin is to employ two stages of laser cooling with 399\,nm and 556\,nm light, where the latter exploits  the $(6s^{2})$~$^{1}S_{0}$ $-$  $(6s6p)$\,$^{3}P_{1}$  inter-combination line.  The natural linewidth of which is 184\,kHz and has a corresponding Doppler cooling limit of 4.4\,\uK.  To generate frequency-stable 556\,nm light a method often employed is  stabilization by locking to a high-$Q$ optical cavity~\cite{Kuw1999,Mar2003b}.  Rather than take this approach, and despite the low scattering rate associated with the \ICL, we employ   sub-Doppler spectroscopy upon a thermal Yb beam that leads to a Zeeman slower and magneto-optical trap.  
In doing so, we have identified an inverted Lamb dip with the  $^{1}S_{0\, F=1/2} -\, ^{3}P_{1\,F'=3/2} $  and $^{1}S_{0\,F=1/2} - \,^{3}P_{1\, F'=1/2}$ 
hyperfine transitions in \Yb.   This was first noted recently in~\cite{McF2016}, where an emphasis was made in relation to its applicability to laser cooling of \Yb. 
Here  we provide a detailed description of the nature of the resonance,
 but also  show the suitability of the $^{1}S_{0\,F=1/2} - \,^{3}P_{1\, F'=1/2}$   line as an effective frequency reference.   For the two hyperfine lines the application of a dc magnetic field transforms a two-level scheme into a four-level scheme, with degeneracy only marginally lifted in the ground state.  The atoms are excited through $\pi$-transitions, while the de-excitation process completes a simple optical pumping scheme.  The enhanced absorption is as an inverted crossover resonance (ICR), where the principal lines involved in the crossover are transitions to symmetrically separated Zeeman substates.  In the case of  \Yb$_{F'=1/2}$\footnote{To distinguish the two \Yb\ \coolingT\ hyperfine lines we will only state the upper $F$ value for the sake of simplicity.} this appears to  be the first instance of an inverted crossover resonance where $\Delta F=0$.  While the ICR for  \Yb$_{F'=3/2}$  is successfully used to stabilize the frequency for laser cooling, the \Yb$_{F'=1/2}$ line also acts as a simple and reliable frequency reference with statistical and systematic uncertainties in the 10\,kHz range.  
In contrast to an ordinary crossover resonance where the signal is limited to the summed strength of the principal lines, the ICR magnitude can be sizeably larger than this factor of two, since the saturation behaviour of  (closed) two-level systems no longer applies.
We note that such resonances are rarely observed on atomic beams because  the natural linewidth of the transition needs to be below a few megahertz (for a sufficiently collimated atomic beam).   By carrying our similar measurements with $^{173}$Yb and detecting  extremely weak ICRs, we find that such ICRs are most apparent when  $I=1/2$. 

  In this paper we open with a description of the sub-Doppler spectroscopy on the Yb intercombination line and commence with an example of an ordinary crossover resonance, before showing results relating to, and describing, the \ICR.  We develop a four-level rate equation model that adequately describes the behaviour seen experimentally.  
   Section~\ref{fstab} describes how the ICR is used for laser frequency stabilization, along with details about absolute frequency measurements  of the  \Yb$_{F'=1/2}$ line.  We complete the discussion with a test of temperature versus frequency instability in a dual-colour MOT using the \Yb$_{F'=3/2}$ line  and the ICR.

\section{Sub-Doppler spectroscopy}  

\begin{figure}[h]			
 \begin{center}
\resizebox{0.6\textwidth}{!}  
{		
  \includegraphics[width=6cm,keepaspectratio=true]{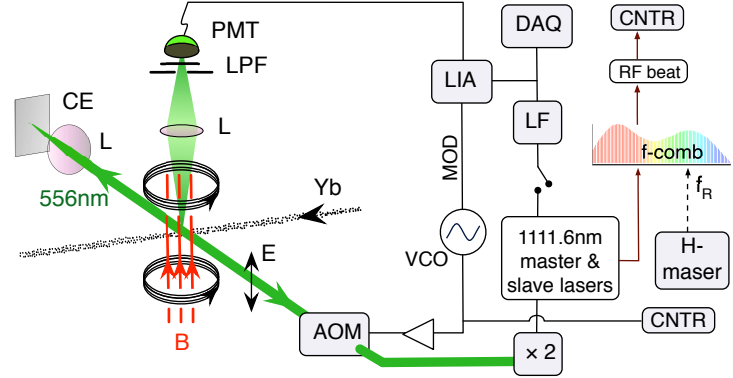}}   %
\caption[]{\footnotesize     %
 Sub-Doppler frequency modulated spectroscopy of the \coolingT\ transition in an atomic beam of Yb. 
   There is also a concave mirror  below the interaction zone redirecting scattered light towards the PMT. 
 AOM, \AOM; DAQ, data acquisition;  $\vec{B}$, magnetic field; CE, cat's eye reflector; CNTR, frequency counter; $\vec{E}$, electric field plane of polarization;  $f_{R}$, frequency comb's mode spacing; L, lens; LF, loop filter (servo gain); LIA, lock-in amplifier; LPF, long-wavelength pass filter (505\,nm); MOD, modulation input; PMT, \PMT\ (cell);   VCO, \VCO.    
  }
   \label{SASexp}  %
\end{center}
\end{figure}
%

 The scheme for the sub-Doppler frequency modulation (FM) spectroscopy  is illustrated in Fig.~\ref{SASexp}, where a retro-reflected 556\,nm beam passes perpendicularly through  the ytterbium atomic beam  and the 556\,nm fluorescence is detected by a \PMT. 
 A lens-mirror combination (cat's eye) is used to produce the retro-reflection. 
   The scheme is similar to that used for many \SAS\, schemes~\cite{Bar1969, Han1971a}, however, given that we observe enhanced absorption, rather than saturation, we refer to the scheme as sub-Doppler FM spectroscopy.  
    Frequency modulation at 33\,kHz is applied to the light via an \AOM\ (AOM) allowing for  $n^{th}$ harmonic detection with a lock-in amplifier.   Unless otherwise mentioned $n=3$. 
    Helmholtz coils set the magnetic field in the vertical direction.     The polarization, $\vec{E}$,  of the 556\,nm light is set  either parallel to that of $\vec{B}$ or perpendicularly.  For the cases of enhanced absorption,  $\vec{E}$ is parallel to $\vec{B}$, which is the case for most of the results presented here.  
  The intensity profile of the 556\,nm beam is elliptical with the major axis (4\,mm in diameter) aligned with the atomic beam.     The 556\,nm light is produced via resonant frequency doubling of 1112\,nm light, where a bow-tie type doubling cavity has at its prime focal point an 18\,mm length \ppKTP\ crystal for the nonlinear conversion, and comprises mirrors with 75\,mm radius of curvature.  The 1112\,nm light is sourced from a commercial fibre laser that injection-locks a ridge waveguide diode laser (Eagleyard EYP-RWL-1120) for amplification to \si45\,mW~\cite{Kos2015}.   
  
An effusion cell   heated above 400\degC\ produces a beam of Yb atoms (with natural isotopic abundance) that escape a reservoir through a trio of stainless steel tubes with a diameter of 1\,mm and length 20\,mm. Any accumulation  of Yb within the tubes is likely to increase the aspect ratio and reduce the angular divergence of the atoms.  
No further collimation techniques are performed. 
 The  fluorescence  from the atom-light interaction passes through a spatial filtering scheme  to minimize stray light (along with an edge filter to block scattered 399\,nm light from a Zeeman-slower beam).
   The fast-linewidth of the 556\,nm light we presume to be \si240\,kHz, based on the  manufacturer's linewidth specification for the 1112\,nm fibre laser (thus, slightly larger than the natural linewidth of the \coolingT\ line). 
   Frequency tuning of the 556\,nm light is carried out via a piezo transducer in the 1112\,nm laser. We calibrate the frequency tuning by use of a frequency comb that is referenced to a hydrogen maser. Further comments regarding the comb are made below when characterizing the frequency instability of the light locked to the ICR.

Crossover resonances are commonly observed in \SAS.  Inverted crossover resonances are less frequently observed, but their occurrence has been known for several decades~\cite{Han1971,Pap1980,Nak1997}.  The element that is unique here is that the inverted crossover resonance arises from the Zeeman splitting of a $\Delta F=0$ transition.  Crossover resonances are most easily observed in vapour cells where there is a full range of velocity components.  Atomic beams, on the other hand, have a very restricted range of transverse velocity; hence the possibility of observing crossover resonances is very much reduced. 
\begin{figure}[h]
 \begin{center}
\resizebox{0.7\textwidth}{!}  %
{		
  \includegraphics[width=7.0cm,keepaspectratio=true]{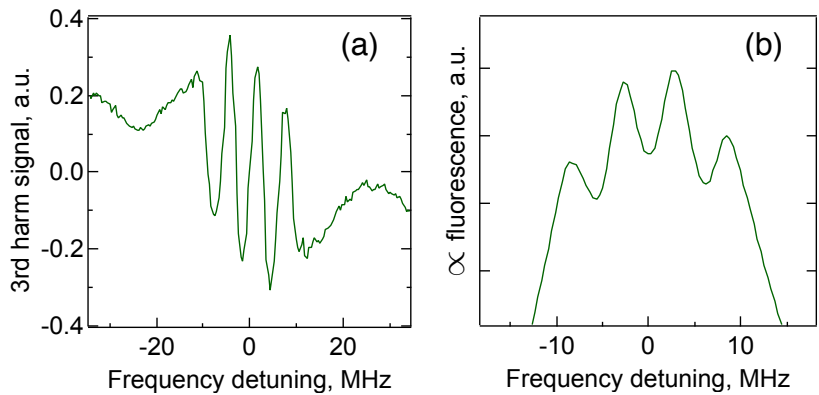}}   %
\caption[]{\footnotesize     %
Saturated fluorescence spectroscopy of the \coolingT\ line in a beam of \Ybtwo\ atoms.  The magnetic field is 0.28\,mT. (a) Third harmonic signal generated by a lock-in amplifier and integrating this yields (b).    The dip in the centre of (b) is the ordinary crossover resonance and the dips to either side are the  Zeeman lines.    $\vec{E}$ is perpendicular to $\vec{B}$ to permit the driving of $\pm\sigma$ transitions.   
 }
   \label{CrossoverResult}  %
\end{center}
\end{figure}

\begin{figure}[h]
 \begin{center}
\resizebox{0.6\textwidth}{!}  %
{		
  \includegraphics[width=9.0cm,keepaspectratio=true]{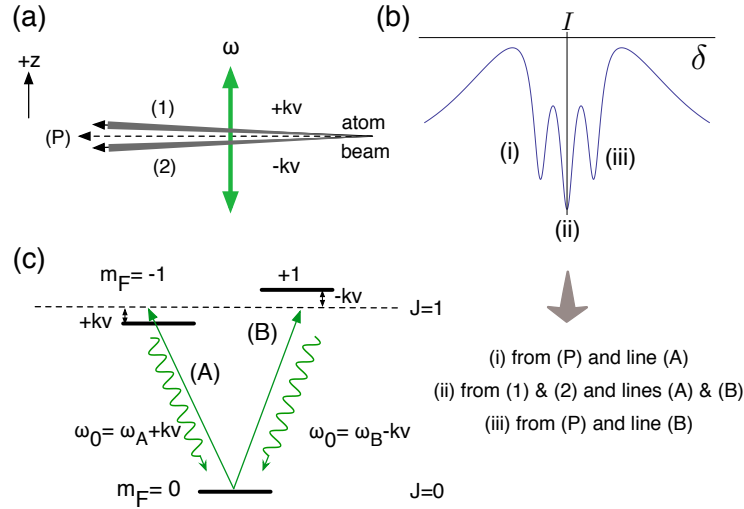}}   %
\caption[]{\footnotesize     %
Description of the crossover resonance in relation to a thermal beam and Zeeman splitting. 
This situation occurs when driving bosonic Yb atoms with $\vec{E}\bot \vec{B}$.  Panel (a) portrays the atomic beam and (velocity) subsets  within the beam that contribute to the fluorescence signals shown in (b), where $I$ denotes intensity and $\delta$ the frequency detuning.  The absorption features are identified with labels (i) to (iii) and the origin of each is summarized.  (P) represents the atoms flowing perpendicularly to the 556\,nm laser beams, identified by frequency $\omega$.  (\,c) shows the transitions responsible for the resonances.  Further details are provided in the text. 
 }
   \label{CRexplanation}  %
\end{center}
\end{figure}

Crossover resonances preserving the saturated absorption can be produced with the bosonic ytterbium isotopes (even mass number) when the excitation occurs through $\sigma$-transitions.  This is made possible when $\vec{E}$ and $\vec{B}$ are orthogonal to each other (i.e., with linear-$\sigma$ polarization\cite{Gry2010b}.  An example is shown in Fig.~\ref{CrossoverResult} taken for \Ybtwo\ and the \coolingT\ line. Here the nuclear spin $I=0$  (for  isotopic and hyperfine line frequency separations for the intercombination line in ytterbium see~\cite{Cla1979,Wij1994,Pan2009,McF2016}).   
Fig.~\ref{CrossoverResult}(a) is the third harmonic output of the \LIA\ and Fig.~\ref{CrossoverResult}(b) is the corresponding absorption signal (produced by integrating (a)).   The outer dips are produced by the $J=0,\, m_{J}=0 \rightarrow J'=1,\, m_{J}=\pm1$ transitions, and the central dip is the crossover resonance.   An explanation for the line spectrum  is summarized in Fig.~\ref{CRexplanation}.    As is commonly observed, the crossover resonance is stronger than the principal lines because there are two groups of atoms contributing to the signal, as opposed to one.  This is portrayed in the upper left panel of Fig. 3.  The principal lines are those generated by the atoms marked $P$ (those flowing perpendicularly to the 556\,nm laser beams), and the crossover resonance is produced by the combined subsets of (1) and (2).  Note, this image should not be taken too literally, because the subsets are in velocity space, while the image is drawn in a spatial sense.  The transitions in this case are closed, having no alternative pathways. In the measurement of Fig. 2 the 556\,nm combined beam intensity was \si160\,$I_{\mathrm{sat}}$  (where the saturation intensity $I_{\mathrm{sat}}$ is 1.4\,W\,m$^{-2}$), thus causing artificial broadening of the lines. Also affecting the resonance widths is the strength of the modulation applied to the 556\,nm light. This is discussed more below.   
The Zeeman line separation, here   at 0.28\,mT, is consistent with the first order Zeeman shift of 21\,MHz$\cdot$mT$^{-1}$ for the $m_{J}=\pm1$ substates (which applies to all the bosonic isotopes).

 \begin{figure}[h]
 \begin{center}
\resizebox{0.5\textwidth}{!}  %
{		
  \includegraphics[width=7.0cm,keepaspectratio=true]{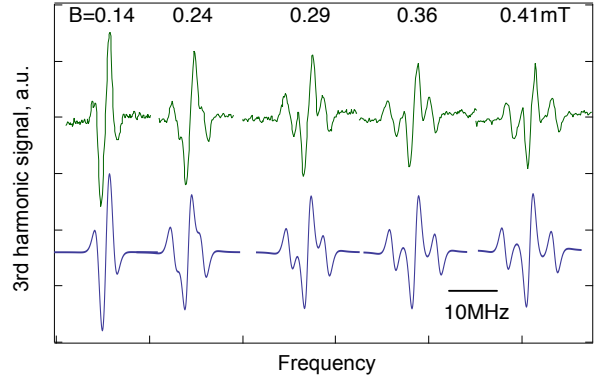}}   %
\caption[]{\footnotesize     %
The \Yb$_{F'=1/2}$ inverted crossover resonance for  a range of magnetic fields with the use of third harmonic detection.    Upper traces: experimental data. Lower traces:  simple model with 14\,MHz$\cdot$mT$^{-1}$ Zeeman shift.    
 }
   \label{DataversusModel}  %
\end{center}
\end{figure}
   
   \begin{figure}[h]
 \begin{center}
\resizebox{0.7\textwidth}{!}  %
{		
  \includegraphics[width=9cm,keepaspectratio=true]{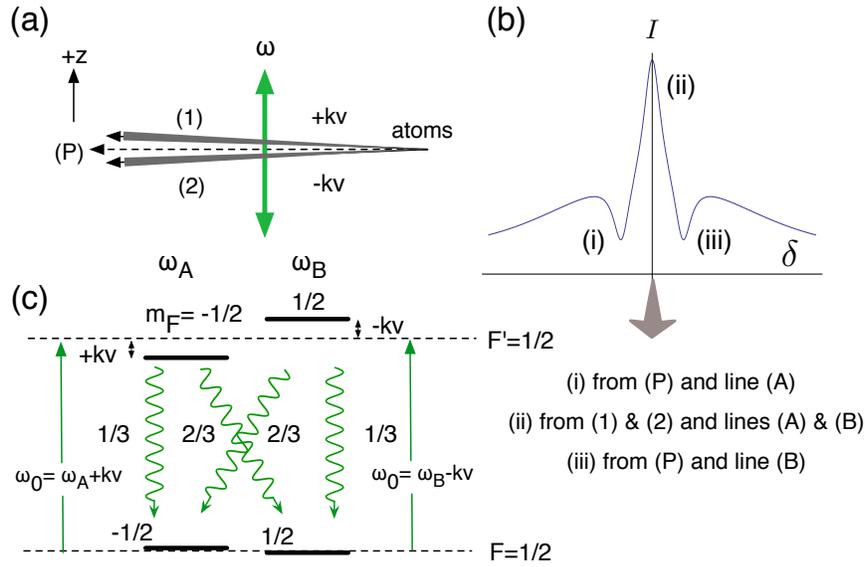}}   %
\caption[]{\footnotesize     %
Description of the inverted crossover resonance when a small bias $B$-field is applied and with $\vec{E}$ $\parallel$ $\vec{B}$. Panel (a) depicts three (velocity) subsets of atoms within a slowly diverging atomic beam  that contribute to the fluorescence signals shown in (b).  
Panel © shows the relevant energy levels. The branching ratios from the upper state Zeeman sub-levels in \Yb\ for the $^{1}S_{0\,F=1/2}-\,^{3}P_{1\,F'=1/2}$  transition are indicated.
   Since the 556\,nm light's polarization is parallel to the magnetic field, $\Delta m_{F}=0$ for the excitation phase. The laser frequency detuning $\delta=\omega-\omega_{0}$, where $\omega_{0}$ is the frequency separation of the states at zero $B$-field. 
 The Zeeman shifts are: 14\,MHz$\cdot$mT$^{-1}$ for the upper $F'=1/2$ state and 3.76\,kHz$\cdot$mT$^{-1}$ for the ground state.
 }
   \label{ICRexplanation}  %
\end{center}
\end{figure}

\section{Inverted crossover resonance: measurement and modelling}  %

The same experimental arrangement was applied to the \Yb\ atoms (with $I=1/2$), except with $\vec{E}$ parallel to $\vec{B}$, ensuring that only $\pi$-excitations occur.   A series of spectra are shown in Fig.~\ref{DataversusModel} for  
the  $^{1}S_{0\,F=1/2} - \,^{3}P_{1\, F'=1/2}$  line with each trace recorded for a different bias magnetic field. The upper traces are experimental data and the lower traces are lineshape models. A similar sequence occurs when the direction of the bias is reversed (for the case of zero field, see~\cite{McF2016}).    One first notices the Zeeman components with $B$  at 0.24\,mT and at slightly higher fields they are easily resolved.  Note the opposing slopes of the main discriminator and the Zeeman components, indicating that the crossover resonance is inverted.  The mean width across the main discriminator for this set of measurements   is \si1.7\,MHz, which is dominated by intensity broadening $-$ the intensity having been set to 90\,$I_{\mathrm{sat}}$,   and the effect of strong frequency modulation. 
The applied modulation amplitude was 2\,MHz. 
 At zero $B$-field there can be a weak saturated absorption signal arising because the four-level structure collapses into a two-level scheme.   This is more readily observed at higher intensities.   At low intensities (\si50\,$I_{\mathrm{sat}}$) the ICR may remain, but is very much  suppressed.  
 At \si0.12\,mT the ICR reaches its maximum contrast and from here the sequence in Fig.~\ref{DataversusModel} follows.

  Before giving a fuller description, the resonances can be  modelled in a basic way with a sum of Lorenztian lineshapes: two separate saturated lines and a central inverted line. The Lorenztian function has the form:  $f_L(\delta,b)\propto(1+(2(\delta-b)/\Delta)^2)^{-1}$, where $\delta$ is the  frequency detuning, $\Delta$ is the  full-width-half-maxium and $b=g_F m_F \mu_B B/\hbar$ we denote for the Zeeman shift of the magnetic sublevels, with $g_{F}$ the Land\'e g-factor and $\mu_B$ the Bohr magneton.   Representing the signal from the lock-in amplifier with third harmonic detection we have,
\begin{equation}
\label{ }
S_L(\delta)\propto\frac{\mathrm{d}^{3}}{\mathrm{d}\delta^{3}}[-f_L(\delta,-b)+ \mathcal{R} f_L(\delta,0)-f_L(\delta,b)]
\end{equation}
 The ratio, $\mathcal{R}$, between the strength of the inverted line to the principal (Zeeman) lines is treated as a free parameter here.  Such an assumption is not used below in a more complete description of the four-level system. 
   For both the experimental data and the summed Lorentzian model the third derivative is shown in Fig.~\ref{DataversusModel}. One sees a clear correspondence between the two sets of traces, with similar features appearing as the Zeeman components separate.   For the $^{171}$Yb$_{F'=1/2}$  line here $g_{F}=28$\,MHz$\cdot$mT$^{-1}$, and $\mathcal{R}=2.7$, notably larger than two.    

  A partial explanation for the inverted crossover resonance is seen in Fig.~\ref{ICRexplanation}.    A key aspect is the four-level structure and  two branching routes for each upper Zeeman level (the ground state Zeeman splitting is $2.7\times10^{-4}$ times weaker than that of the upper state splitting). 
  The branching ratios from the upper states are readily calculated for the $^{171}$Yb$_{F'=1/2}$  and $^{171}$Yb$_{F'=3/2}$ systems~\cite{McC1996}, given that mixing with the $^{1}P_{1}$ state permits the \coolingT\ transition to occur.   Fig.~\ref{ICRexplanation} shows the branching ratio for the $^{1}S_{0\,F=1/2} - \,^{3}P_{1\, F'=1/2}$  hyperfine line.   
 Two subsets of atoms combine to produce the ICR, each with the same absolute value of Doppler shift ($kv$), but opposing sign. 
   For the excitation phase the 556\,nm light polarization is parallel to the magnetic field, implying $\Delta m_{F}=0$.
 When the optical frequency $\omega_{0}, $ lies midway between the upper state separation then both subsets of atoms participate maximally in the absorption and cycling process (and $\omega_{0}$ is the mean of $\omega_{A}$ and $\omega_{B}$).   Atoms (1) with $+kv$ Doppler shift interact with the $-z$  beam, exciting transition (B) such that $\omega_{B}=\omega_{0}+kv$. Alternatively, these atoms can interact with the $+z$  beam, exciting transition (A), where $\omega_{A}=\omega_{0}-kv$.   Whichever decay channel occurs there is always the possibility for re-excitation, hence  atoms are not lost to a dark state. 
 The situation is reversed for atom subset (2).  
 The width of the dispersive curve reduces to  \si800\,kHz when the 556\,nm beam intensity is 30\,$I_{\mathrm{sat}}$ (there are other factors influencing the width that are discussed below). 
  For the most probable velocity of 330\,m\,s$^{-1}$, this corresponds to 1.4\,mrad of separation between the two diverging sets of atoms responsible for the \ICR.   

  \begin{figure}[h]
 \begin{center}
\resizebox{0.8\textwidth}{!}  %
{		
  \includegraphics[width=8.0cm,keepaspectratio=true]{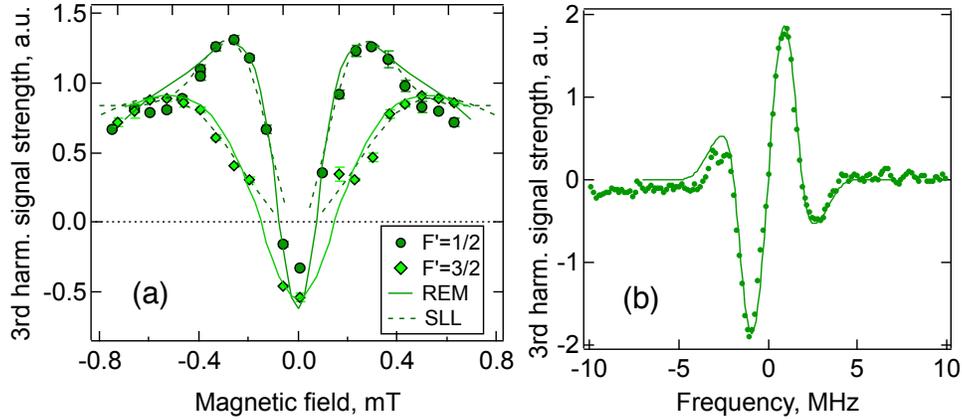}}   %
\caption[]{\footnotesize     %
(a) Signal strength from the sub-Doppler spectroscopy versus magnetic field for the two $^{171}$Yb hyperfine lines.  At zero $B$-field the magnetic sublevels remain degenerate and  saturated absorption occurs (represented by the inverse signal strength).  REM refers to outputs generated with the four-level  rate equation model and SLL refers to a sum of Lorentzian line shapes (further described in the text). 
(b)~The spectrum for $^{1}S_{0\,F=1/2} - \,^{3}P_{1\, F'=1/2}$ line with third harmonic detection and  $B=0.15$\,mT.  Overlaid is a curve generated with the rate equation model using experimental parameters ($s_0=88$). 
 }
   \label{Symmetry}  %
\end{center}
\end{figure}

The strength of the resonance exhibits symmetrical behaviour for opposing magnetic field directions,
as shown in Fig.~\ref{Symmetry}(a).   The vertical axis represents the size of the discriminator from the output of the \LIA,  again recorded with 3$^{rd}$ harmonic detection.  At zero $B$-field the Zeeman sublevels remain degenerate, thus acting like a two-level system and the usual saturated absorption behaviour occurs. Applying the $B$-field in either direction causes the Zeeman level splitting  and  two decay channels to occur for each excited Zeeman state.  
The situation is similar for the $^{1}S_{0\,F=1/2} - \,^{3}P_{1\, F'=3/2}$   
  line (there is a frequency separation of 5937.78\,MHz between the two \Yb\ hyperfine lines); however, there is a reversal of branching ratios where the $\sigma$ channel de-excitation occurs with $1/3$ probability and the $\pi$ channel de-excitation occurs with $2/3$ probability.  
The maximum signal strength and contrast is greater for the $F=1/2\rightarrow1/2$ line than the  $F=1/2\rightarrow 3/2$ line. 
  The reduction in signal strength at higher $B$-field takes place as the two Zeeman lines separate (also seen in Fig~\ref{DataversusModel}).   The Zeeman shift of 14\,MHz$\cdot$mT$^{-1}$ for $F=1/2\rightarrow 1/2$ is twice as strong as that for 
 $F=1/2\rightarrow 3/2$; hence the lines become resolved at lower fields for the former and the main discriminator strength falls away more rapidly with $B$-field.  The dashed lined curve-fits away from saturation (of Fig.~\ref{Symmetry}a) are derived from the simple summed-Lorentzian model described above (and cannot reproduce saturation). 
 The abscissa scaling  between the $F'=1/2$ and  $F'=3/2$ data fits precisely the  factor of two ratio of the Zeeman shifts.  
The peak-to-peak values versus magnetic field in Fig.~\ref{DataversusModel} has a slightly different trend compared to the data in  Fig.~\ref{Symmetry}(a), \wrt\ the $F'=1/2$ data.    The main reason being  that the intensity used for Fig.~\ref{Symmetry}(a) was \si180\,$I_{\mathrm{sat}}$, thus broadening the lines and influencing the progression seen in Fig.~\ref{DataversusModel} (the frequency modulation amplitude  was also twice as strong).  

To encompass the full range of behaviour of the four level system a rate equation model was developed to make comparisons with the experimental data.  
With a 4\,mm laser beam width (in the direction of the flow of atoms) and a most probable velocity of 330\,m\,s$^{-1}$, the atomic transit time across the beam is estimated to be about 12\,$\mu$s, which is large compared to the lifetime of the $^3P_1$ level (0.86 $\mu$s).  To a good approximation the atomic evolution should reach the steady-state, even at high saturation. 
  In this case, the result of the laser-atom interaction can be estimated by the result of a rate-equation model.

\begin{figure}[h]
 \begin{center}
\resizebox{0.8\textwidth}{!}  %
{		
  \includegraphics[width=8cm,keepaspectratio=true]{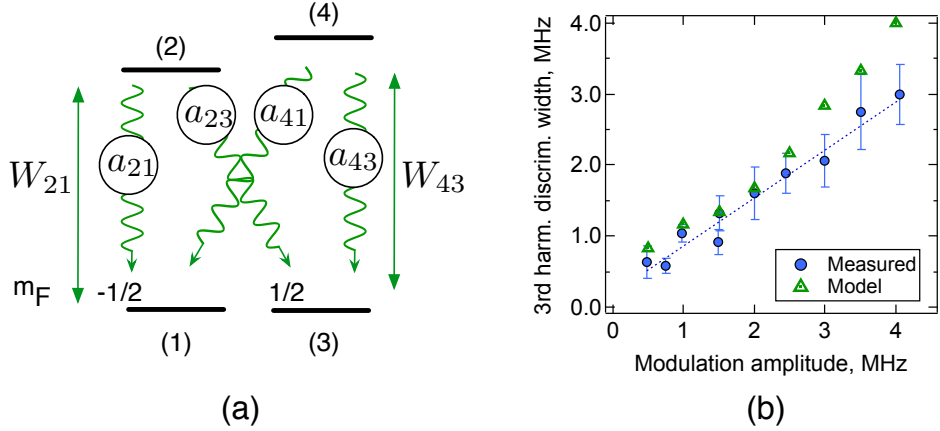}}   %
\caption[]{\footnotesize     %
(a) Relevant energy levels and transitions for modelling the \coolingT\  $^{171}$Yb inverted crossover resonance.  The same representation can be used for the two hyperfine lines ($F'=1/2$ and $F'=3/2$); only the branching coefficients change.  
(b)~The third harmonic discriminator width of the ICR spectrum ($F'=1/2$) versus  modulation amplitude: $s_{0}=90$ and $B=0.1$\,mT.  
 }
   \label{ModelandMod}  %
\end{center}
\end{figure}

A set of rate equations is solved describing  population changes in a four-level system, where the fractional population of the four levels is $n_{j}$, $j\in$ \{1,4\}.  
 The labelling of the states is shown in Fig.~\ref{ModelandMod}(a).   
In steady state, the rate equations for a $\pi$ polarized light become,
\begin{equation}
\eqalign{
n_{1}(W_{12}+\gamma_t) -\gamma_t n^0_1-n_{2}(W_{12}+a_{21}\Gamma)-n_{4}a_{41}\Gamma =0 \cr
n_{2}(W_{12}+\Gamma+\gamma_t)- n_{1}W_{12}=0 \cr
n_{3}(W_{34}+\gamma_t) -\gamma_t n^0_3 -n_{4}(W_{34}+a_{43}\Gamma)-n_{2}a_{23}\Gamma =0 \cr
n_{4}(W_{34}+\Gamma+\gamma_t)-n_{3}W_{34}=0 }
\label{eqsteadystate}
\end{equation}
where $\Gamma$ is the spontaneous rate of emission, 
 $W_{ij}$ is the pumping rate of the transition with levels $\{i,j\}$, $a_{ij}$ are the relaxation branching coefficients and $\gamma_{t}$ is a relaxation term introduced to account for the finite interaction time as the atoms  transit  the laser beams.   The ground states will have equal population before the interaction (denoted $n_{i}^{0}$) and without any loss of atoms the sum $\sum_{i=1}^4 n_{i}=1$ holds.   The solutions for  $n_i$ are given in Eq.(\ref{nsolutions}) in the Appendix.


The driving rates are found from,
\begin{eqnarray} 
\eqalign{
\fl W_{12}=a_{21} \sigma_{0}s_{0}N_{S}\cr
 \times  \left(\frac{\Gamma_2\Gamma/4}{[\delta_{12}-kv+a_m \cos(x)]^{2}+(\Gamma_{2}/2)^{2}}+ \frac{\Gamma_2\Gamma/4}{[\delta_{12}+kv+a_m \cos(x)]^{2}+(\Gamma_{2}/2)^{2}}\right),\cr	
\fl W_{34}=a_{43} \sigma_{0}s_{0}N_{S} \cr
 \times \left(\frac{\Gamma_2\Gamma/4}{[\delta_{34}-kv+a_m \cos(x)]^{2}+(\Gamma_{2}/2)^{2}}+\frac{\Gamma_2\Gamma/4}{[\delta_{34}+kv+a_m \cos(x)]^{2}+(\Gamma_{2}/2)^{2}}\right)
}
\label{pumprates}
\end{eqnarray}
where $s_{0}$ is the on-resonance saturation parameter, $\sigma_{0}=3\lambda^{2}/2\pi$ is the  light-atom  scattering cross section (on resonance),  $k$ is the wave number and $\Gamma_{2}$ is the combined line-width associated with spontaneous emission, transit time and laser frequency noise, $\Gamma_{2}=\Gamma+\gamma_t+\gamma_L$, with $\gamma_L$ the laser linewidth.
$a_m \cos(x)$ is the amplitude modulation term which will be defined more completely below. 
 $N_{S}=I_{\mathrm{sat}}/\hbar\omega_{0}$ represents the number of photons per unit area at the saturation intensity. The frequency detuning terms account for the Zeeman shift  through, 
\begin{equation}
\label{ }
\delta_{ij}=\delta +( m_i g_{i}- m_{j}g_{j})B \mu_B/\hbar
\end{equation}
where $\delta=\omega-\omega_{0}$ is the laser frequency detuning from the centre of resonance in null magnetic field, $\omega_{0}$, $g_{i}$ is the Land\'e factor of the level $i$, and $\hbar$ is the reduced Planck constant. We note the ground state splitting is negligible in comparison to the upper state splitting. 
The laser frequency is sinusoidally modulated and the fluorescence signal  is detected by means of  third harmonic output of a lock-in amplifier. The modulation frequency, $\omega_m $, is fast in comparison to the mean laser-frequency scanning  rate, i.e. the detuning, $\delta$, is approximately constant during a modulation period, so that we can insert the modulation term in the steady-state equations. The $a_m$ term is the frequency modulation amplitude (in rad$\cdot$s$^{-1}$) 
and $x=\omega_m t$ is the modulation term's phase with $t$ the time.   
In this case, following from \cite{Wah1961,Wil1963,Isl1969}, the $q$-th Fourier-series coefficient of the modulated signal forms the signal yielded by the lock-in detection employing in-phase detection at the $q$-th harmonic of the modulation frequency. The combined signal  is evaluated through,  
\begin{equation}
S_{q}(\delta) = \frac{2 \Gamma}{\pi}\int_{-v_{t}}^{v_{t}}\int_{0}^{\pi}\{n_{2}(v,\delta,x)+n_{4}(v,\delta,x)\} \cos(q x)\,f(v)\mathrm{d}x\mathrm{d}v
\label{eq:pop}
\end{equation}
where the rate equations are solved in a stepwise manner assuming a top-hat transverse velocity distribution $f(v)$ truncated at $-v_{t}$ and $v_{t}$.  The equation incorporates an evaluation of the photon fluorescence rate produced by atoms in the two upper states. 
We note that Eq. (\ref{eq:pop}) remains valid if the modulation amplitude is not large in comparison to the intensity broadened linewidth.    
 Figure~\ref{ModelandMod}(b) provides an example of resonance width versus  frequency modulation amplitude at $s_0=90$ and $B=0.1$\,mT for both measurement and model.  The two agree reasonably well until a modulation amplitude of \si2.5\,MHz is reached.  We choose $v_t=10$\,m\,s$^{-1}$, which is well beyond the Doppler components that compensate for the Zeeman shifts explored here.  
 
  The relaxation branching ratios of Eq.~\ref{eqsteadystate} can be summarised with $a_{21}=1-a_{23}=a_{43}$ and $a_{41}=a_{23}$.  For $^{1}S_{0\,F=1/2} - \,^{3}P_{1\, F'=1/2}$, $a_{23}=2/3$ and for $^{1}S_{0\,F=1/2} - \,^{3}P_{1\, F'=3/2}$, $a_{23}=1/3$.  These branching ratios may not be exact, since the intercombination line is a spin-flip transition (that is permitted through mixing with the $^{1}P_{1}$ state). 
Fig.~\ref{Symmetry}(b)  shows the spectrum for $^{1}S_{0\,F=1/2} - \,^{3}P_{1\, F'=1/2}$ line with third harmonic detection, $s_0=88$,  $B=0.15$\,mT and a frequency modulation amplitude of 2\,MHz.   The solid (green) line is generated by the rate equation model for the same magnetic field, intensity and modulation amplitude.  Only vertical scaling is applied to the computed result.  The agreement between the experimental data and the model lends credence to its validity. 

 The comparison between model and data is also shown in Fig.~\ref{Symmetry}(a).  The solid lines are the computed signal strengths for the $^{1}S_{0\,F=1/2} - \,^{3}P_{1\, F'=1/2}$ and $^{1}S_{0\,F=1/2} - \,^{3}P_{1\, F'=3/2}$  lines.    The input parameters match those of the experiment, where $s_0=176$ and the frequency modulation amplitude is 4\,MHz. 
The model shows the same characteristic behaviour across the full magnetic field range:  from saturation at and near zero $B$-field, to the reverse peaks away from zero field.  However,   we have applied independent scaling between the data and model for the two hyperfine lines. The model creates different ICR strengths for the two hyperfine lines because of the different branching ratios, but the ratio is not in accord with that observed in the experiment.   In the model the  $F'=3/2$  resonance is \si15\,\% stronger than the $F'=1/2$   ICR.   This difference in ratio may be because the  model does not take into account the spatial dependence of fluorescence.  The quantity of  fluorescence heading in the direction of the PMT differs depending on whether the released radiation is $\pi$ or $\sigma$ polarised~\cite{Gry2010c}.  For the two cases the ratio of $\pi$-to-$\sigma$ polarised light changes depending on the branching ratios. 
 Another possible, but unlikely, factor influencing the behaviour of the $F'=3/2$ line shape  is the presence of the $^{173}$Yb  $^{1}S_{0\,F=5/2} - \,^{3}P_{1\, F'=3/2}$ line, which lies only \si2.6\,MHz away (the $^{173}$Yb  line is approximately one sixth the strength of the $^{171}$Yb$_{F'=3/2}$ line). 

\section{Frequency stabilization} \label{fstab}

\begin{figure}[h]
 \begin{center}
\resizebox{0.8\textwidth}{!}  
{		
  \includegraphics[width=12cm,keepaspectratio=true]{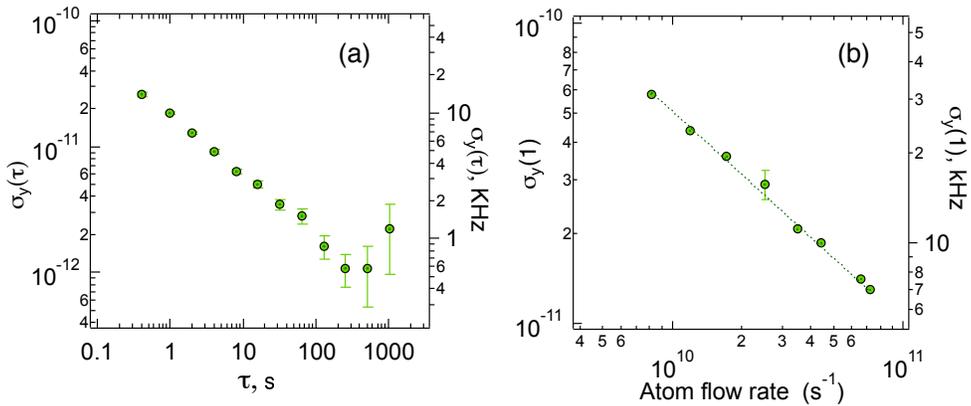}}   %
\caption[]{\footnotesize     %
(a)  Frequency instability of the 556\,nm light when locked to the $^{171}$Yb$_{F'=3/2}$ \ICR\ for an atomic flow rate of $\sim4\times10^{10}\,s^{-1}$.  (b) Frequency instability at $\tau=1$\,s and its dependence on atomic flow rate. 
 }
   \label{Stability}  %
\end{center}
\end{figure}

The ICR with the FM spectroscopy produces a dispersive signal  that one can use to stabilize the 556\,nm laser frequency.  The correction signal is sent to the 1112\,nm fibre laser through a piezo transducer with a servo bandwidth of approximately 20\,Hz (deliberately set low to avoid adding technical noise at higher Fourier frequencies).   We assess the frequency instability by counting the frequency of the beat generated by mixing 1112\,nm light with the adjacent element of a frequency comb (whose mode spacing is referenced to a H-maser signal).  Here we consider first harmonic detection in the FM spectroscopy as it confers approximately a factor of two improvement in the frequency instability compared with third harmonic detection (and is relevant for the laser cooling below).   The frequency instability for an atomic flow rate of $N\sim4\times10^{10}\,s^{-1}$ through the beams is shown in Fig.~\ref{Stability}(a).  This result is for the $^{171}$Yb$_{F'=3/2}$ hyperfine line.  The left ordinate is the instability made fractional with respect to the carrier frequency of the 556\,nm light.  
The instability reduces with a $\tau^{-1/2}$ dependence out to  about 600\,s. 
At 50\,ms, which is the transfer duration for the second stage of the MOT, the instability is 45\,kHz, and is well below the natural linewidth of the $^{3}P_{1}-\,^{1}S_{0}$ line (184\,kHz). 
The atomic flow rate was estimated by recording the dc level of the PMT (registering 556\,nm photons), accounting for the collection efficiency and applying the standard scattering rate  equation (for zero frequency detuning).  The uncertainty for $N$ is \si50\,\%. Based on particle flow rates through a cylindrical tube and accounting for atomic beam divergence in the vertical plane, $N$ should be an order of magnitude greater.  The difference may be explained if  Yb  has accumulated to the sides of  the tubes and reduced the aspect ratio. 


  Further improvements in frequency instability   can be produced by increasing the atomic flow rate as seen in Fig.~\ref{Stability}(b).   A power law fit gives a dependence of $N^{-0.7}$.  
This rate of fall is likely a coincidental combination of noise terms proportional to  $(N^{-2} + b N^{-1})^{1/2}$, with $b$ some constant.   We do not observe a change in frequency instability with changes (by an order of magnitude) in optical intensity,  suggesting that photon shot-noise is not a limiting factor, thus an unidentified noise source  in the detection system contributes to the instability.   
 The falling frequency instability is a useful trait for lasers employed in laser cooling.  Moreover,  by locking to the \coolingT\ line one has immediate access to the frequency detuning (unlike in the case of locking to an optical cavity). 

\begin{figure}[h]
 \begin{center}
\resizebox{1\textwidth}{!}  %
{		
  \includegraphics[width=10cm,keepaspectratio=true]{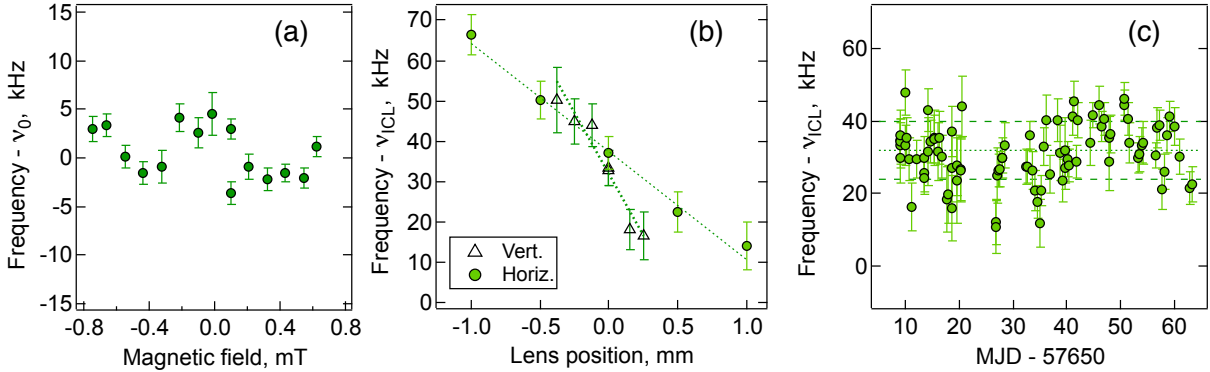}}   %
\caption[]{\footnotesize     %
(a) Optical frequency measurements of the ICR line centre versus magnetic field for \Yb$_{F'=1/2}$.  
(b) The absolute frequency of the \Yb\ $^{1}S_{0\,F=1/2} - \,^{3}P_{1\, F'=1/2}$  hyperfine line frequency versus cat's eye lens position 
with use of the ICR.   Zero corresponds to alignment of the 556\,nm beams with the optical axis of the cat's eye.  (c) Absolute frequency  measurements of the same hyperfine line.  A $B$-field of 0.1\,mT is applied for all measurements to create the ICR.  The offset frequency is $\nu_{\mathrm{ICL}}=539\,384\,469\,000$\,kHz.  
 }
   \label{AbsMesures}  %
\end{center}
\end{figure}

The symmetry of the ICR is again apparent when the line-centre frequency  is measured as a function of magnetic field.   The laser is locked to the centre of the dispersive curve  and frequency counting of the IR beat is made for approximately 100\,s at each $B$-field setting. 
On this occasion we consider the $^{171}$Yb$_{F'=1/2}$ hyperfine line.
 Fig.~\ref{AbsMesures}(a) 
  shows the mean line-centre frequency, $\nu_{F'=1/2}$, versus magnetic field.  The frequency variations are constrained to within $\pm5$\,kHz  across \si1.5\,mT  (or 15\,G).  In fractional terms this is $9\times10^{-12}$, demonstrating that the first order Zeeman shift is efficiently cancelled with this \ICR\ (and the second order dependence is not apparent).   Other systematic shifts have been investigated for the \Yb$_{F'=1/2}$ line.  One can measure the frequency dependence on the separation between the beam and the optical axis of the cat's eye, both in the direction of the atomic beam and in the vertical direction.    In both cases there is  a monotonic shift that is linear  over small displacements, as seen in Fig.~\ref{AbsMesures}(b).  In the horizontal plane  the shift is 26\,kHz\,mm$^{-1}$.  Zero displacement is made evident  by observing the \SNR\ of the signal, or equivalently the Allan deviation of the locked laser frequency.   Alignment between the 556\,nm beams and the optical axis can be determined to within $\pm0.4$\,mm, which equates to an uncertainty of $\pm11$\,kHz.   
   Displacing the lens in the vertical direction creates a linear frequency shift over $\pm0.3$\,mm from the optical axis. A displacement greater than \si0.2\,mm is relatively easily identified and the associated systematic shift is 13\,kHz.   We note that for larger ranges of displacement in the direction  of the atoms the frequency variation can exhibit a quadratic dependence, the curvature of which  
    may depend on the angle between the 556\,nm beams and the optical axis of the cat's eye. 
   Shifts associated with the wavefront curvature were tested by changing the  lens-mirror separation in the cat's eye.  When set to the focal length of the lens the shift is less than 3\,kHz\,mm$^{-1}$.

The AC Stark shift was measured by varying the 556\,nm beam intensity over a range of $s_{0} \in \{10,600\} $ 
 where $s_{0}$ is the on-resonance saturation parameter ($s_{0}=I/I_{\mathrm{sat}}$). 
 The shift is almost indiscernible at $0.034\pm0.019$\,kHz\,W$^{-1}$\,m$^{2}$.   For the absolute frequency measurements the 556\,nm intensity is \si80\,W\,m$^{-2}$ (for the two beams),  
thus the associated correction required  is smaller than the statistical variations between repeated measurements. 
The insensitivity to magnetic field and low sensitivity to other shifts  
prompted the  recording of (semi-) regular measurements of $\nu_{F'=1/2}$ over an extended period of time as a test of the repeatability.  Measurements were taken over a  period of two months and are shown in Fig.~\ref{AbsMesures}(c).  
  The mean optical frequency from this series is $539\,384\,469\,031(8)(23)$\,kHz, where the standard deviation from the weighted mean is  8\,kHz  (which is \si5\,\% of the natural linewidth).   The reduced $\chi^{2}$ for the set is 2.1.  
 Through the quadratic sum of uncertainties associated with systematic shifts (mentioned above) the  total systematic uncertainty is 23\,kHz.  Also included is an uncertainty associated with servo error (more specifically, identifying the centre of the line)~\cite{note1}.   
We note this overall uncertainty is more than an order of magnitude 
smaller than the previous most accurate measurement of a Yb intercombination line frequency~\cite{Nen2016}.    While not at all competitive with cold-atom atomic frequency standards, the \Yb$_{F'=1/2}$  ICR  provides a simple and effective thermal beam frequency reference.   The \Yb$_{F'=1/2}$  line has the advantage that it  is free from neighbouring isotopic lines that can perturb the lineshape, unlike the \Yb$_{F'=3/2}$  line. 
Although calibration of the H-maser was not carried out during this measurement run it is not foreseen to be more inaccurate than \si$5\times10^{-12}$, based on previous calibrations against UTC (coordinated universal time), National Measurement Institute, Sydney.  In support of this, absolute frequency measurements of the \Yb\ \clockTdash\ (clock) transition by use of the H-maser were also recorded in conjunction with some of the \coolingT\ line measurements.  The mean deviation from previously published values of the clock transition frequency  is less than 10\,kHz.  The method used for the clock line frequency measurements was reported in~\cite{Nen2016},  
though notably, we now perform all the the measurements free of light shifts associated with the MOT.    

\begin{figure}[h]
 \begin{center}
\resizebox{0.6\textwidth}{!}  %
{		
  \includegraphics[width=9cm,keepaspectratio=true]{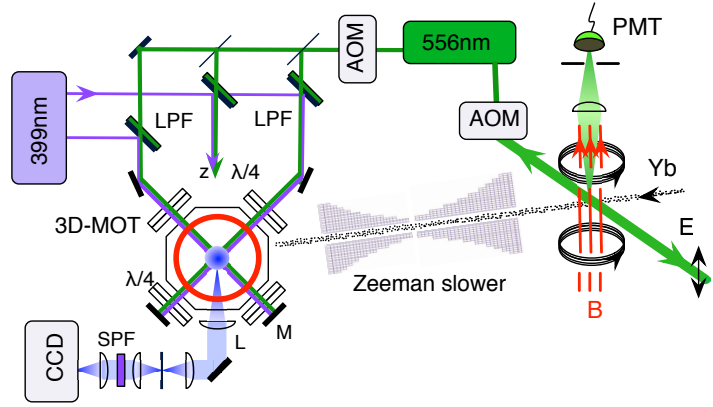}}   %
\caption[]{\footnotesize     %
A dual-colour MOT for ytterbium (using \fastT\ and \coolingT\ lines) with a Zeeman-slower (using the \fastT\ line).  The third axis of the MOT is not shown. The 556\,nm light is frequency stabilized via the inverted crossover resonance produced in the sub-Doppler spectroscopy. Free expansion of the atom-cloud is recorded with the CCD camera.  The $\lambda/4$ refers to  achromatic quarter wave plates suitable for 399\,nm and 556\,nm wavelengths. SPF (LPF), short (long) wavelength pass filters (505\,nm).  Other labels are defined in Fig.~\ref{SASexp}.
 }
   \label{MOT}  %
\end{center}
\end{figure}

\begin{figure}[h]
 \begin{center}
\resizebox{0.8\textwidth}{!}  %
{		
  \includegraphics[width=12cm,keepaspectratio=true]{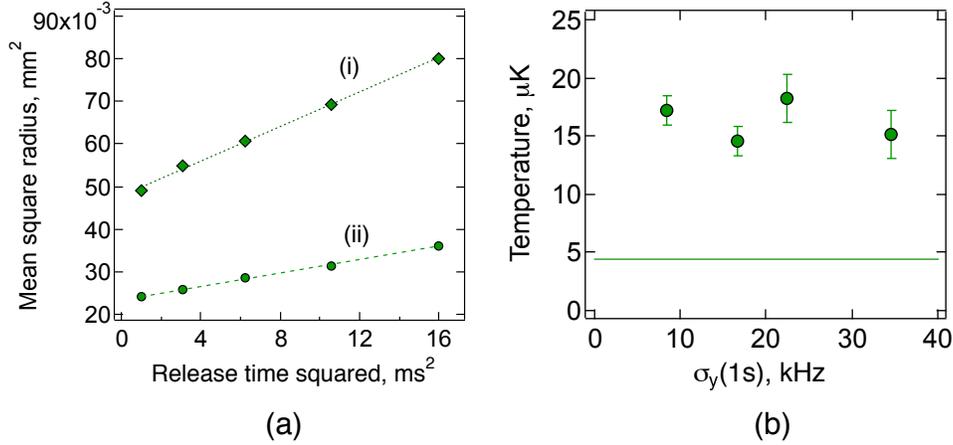}}   %
\caption[]{\footnotesize     %
(a)  Atomic cloud expansion versus time (squared units) for (i) \Ybtwo, and (ii) \Yb, after release from the dual-colour MOT   (b) \Yb\ cloud temperature  for a range of frequency instabilities (at $\tau=1$\,s).  The solid (green) line marks the approximate Doppler cooling limit (for zero intensity). 
 }
   \label{TempvsStability}  %
\end{center}
\end{figure}

In the case of the \Yb$_{F'=3/2}$  ICR, the frequency stabilized 556\,nm light is used to  cool atoms in a dual-colour \MOT.  The experimental set-up is depicted in Fig.~\ref{MOT},  where the sub-Doppler spectroscopy is carried out prior to the Yb atoms reaching a Zeeman slower. The first-stage cooling is performed with 399\,nm light on the \fastT\ transition, then second stage cooling is carried out with 556\,nm light driving the intercombination line. Further details about the \MOT\ are described in~\cite{McF2016}.   In the FM scheme for the laser lock we use first harmonic detection as it grants a higher signal-to-noise ratio and improved frequency instability, by at least a factor of two. 
We performed atomic cloud temperature measurements as a function of the 556\,nm light frequency instability, where the temperature was estimated based on the expansion rate of the cloud and imaging onto a CCD.  Examples of $r^{2}$ versus $t^{2}$ data are shown in Fig.~\ref{TempvsStability}(a) for two isotopes.  The upper (lower) trace is for \Ybtwo\ (\Yb) with a corresponding temperature of 46\,\uK\  (16\,\uK).   The frequency instability was varied by changing the atomic flow rate through varying the oven temperature.  Over the instability  range that could be tested, seen in Fig.\ref{TempvsStability}(b), we do not observe any influence on the cloud temperature, indicating that the remaining frequency fluctuations are not responsible for the temperature limit.     The cause for the limitation  
 is not known,  though we note that the $10-15$\,\uK\ range is a common limit for the cooling of fermionic Yb isotopes; slightly above the Doppler cooling theory  limit of 4.4\,\uK.

\section{Conclusions}  
We have presented results in relation to the recently discovered crossover resonances in neutral ytterbium, notably the inverted crossover resonances (ICRs) produced with the two hyperfine lines in \Yb.   They arise through sub-Doppler spectroscopy of the intercombination line ($\Gamma/2\pi =184$\,kHz) when the atoms are formed into a thermal beam.  This appears to be the first instance where the ICR occurs with $\Delta F=0$ and the principal lines belong to the same Zeeman manifold.    The $I=1/2$ isotope creates a circumstance where only two narrow velocity groups are involved the generation of the resonance. The ICR is apparent when the separation of the principal lines ranges from \si1\,MHz to tens of megahertz, which is unlike the case of ICRs in gas cells, where the separation is typically hundreds of MHz~\cite{Sch1994}. There is a high degree of symmetry in the line profiles due to the nature of the phenomenon and this is confirmed by use of a four-level rate equation model.   The enhanced absorption is particularly striking in the case of the \Yb\ $^{1}S_{0\,F=1/2}\, - \,^{3}P_{1\, F'=1/2}$ hyperfine line. 
By frequency  locking to the centre of the ICR, one can generate a frequency reference that is independent of the Zeeman shift to first order.  
We observe fractional variations less than $9\times 10^{-12}$ over a magnetic field range of $\pm0.8$\,mT.    Hence, the ICR for  \Yb$_{F'=1/2}$ gives a very good representation of line-centre at zero $B$-field.    Absolute frequency measurements of the  \Yb$_{F'=1/2}$  line against a hydrogen maser  were recorded over a two month period.   
 The mean frequency is 539\,384\,469\,031(8)(23)\,kHz. 
   The combined  $1\sigma$ uncertainty of 24\,kHz in fractional terms is $4.5\times 10^{-11}$.  We have investigated relevant systematic shifts, the most significant being the overlap of the forward and return beams used in the sub-Doppler spectroscopy, and errors associated with identifying line-centre.   
   The accuracy of the hydrogen maser was verified by carrying out  \clockT\ line frequency measurements in \Yb.
    We believe this to be  the most accurate Yb intercombination line frequency measured to date. 
The reproducibility of the line frequencies points to  
 a modestly accurate frequency reference that is simple and with the potential to be compact and robust~\cite{Man2012}.   We note that with the atomic flow rates estimated here, 1\,g of Yb would last over one hundred years with continuous operation.   The \Yb$_{F'=1/2}$ frequency can be used to compute a value for the \Yb$_{F'=3/2}$  frequency based on the hyperfine frequency separation of 5\,937\,779\,(57)\,kHz~\cite{McF2016}.  The result is 539\,390\,406\,810\,(62)\,kHz, which is consistent with that previously reported and with reduced uncertainty~\cite{Nen2016}.

For the \Yb$_{F'=3/2}$  hyperfine line, the ICR was used as a means of laser frequency stabilization for laser cooling.   By locking to the centre of the crossover resonance we show that temperatures of sub-20\,\uK\ can be produced in a dual-colour \MOT.   By comparing the atom-cloud temperature to various levels of frequency instability, we find that the remaining instability of the 556\,nm light is not responsible  for the temperature limit; hence locking to the ICR is an effective means of laser stabilization for laser cooling. 
Similar inverted crossover resonances with an atomic beam are expected to be manifest whenever  the species has a nuclear spin of 1/2 and the electronic transition has a sufficiently narrow natural linewidth; other examples include the intercombination  lines in $^{199}$Hg ($\Gamma/2\pi =1.3$\,MHz),  $^{111}$Cd and $^{113}$Cd ($\Gamma/2\pi =65$\,kHz).  

\section*{Acknowledgement}
This work was supported by the \ARC\ (grant LE110100054).  J.M. was supported through an ARC Future Fellowship (FT110100392).  
We thank members of the ARC CoE Engineering Quantum Systems and OBEL for the use of diagnostic equipment, and Nils Nemitz, Fr\'ed\'eric du Burck and Stefan~Weyers  for valuable discussions.




%
  \section*{Appendix: Populations at the steady-state } 
\label{appendix1}
The solutions to the steady state four-level rate equation model adopted here are the following.  The parameters are described in the main text.
\begin{eqnarray}
\eqalign{
\fl n_1=\gamma_t  (W_{12}+\Gamma+\gamma _t ) \left( n_3^0  a_{41} W_{34}\Gamma+n_1^0  \left[ W_{34}( \Gamma(1-a_{43}) +2 \gamma_t )+\gamma_t  (\Gamma+\gamma_t )\right] \right)/D \cr
\fl n_2=\gamma_t  W_{12} (  n_3^0 a_{41} W_{34}\Gamma + n_{10} \left[ W_{34} ( \Gamma(1-a_{43}) +2 \gamma_t  )+\gamma_t  (\Gamma+\gamma_t )\right] )/D \cr
\fl n_3=\gamma_t  (W_{34}+\Gamma+\gamma _t ) \left( n_1^0  a_{23} W_{12}\Gamma+n_3^0  \left[ W_{12}( \Gamma(1-a_{21}) +2 \gamma_t )+\gamma_t  (\Gamma+\gamma_t )\right] \right)/D \cr
\fl n_4 = \gamma_t  W_{34} (  n_1^0 a_{23} W_{12}\Gamma + n_{30} \left[ W_{12} ( \Gamma(1-a_{21}) +2 \gamma_t  ) + \gamma_t (\Gamma+\gamma_t )\right] )/D \cr
\fl D= \gamma_t  (\Gamma+\gamma_t ) \left[ W_{34} ( \Gamma (1-a_{43}) +2 \gamma_t)+\gamma_t  (\Gamma+\gamma_t )\right]
 + W_{12} \{  \gamma_t  (\gamma_t +\Gamma ) (\Gamma (1-a_{21}) +2 \gamma_t )  \cr
 + W_{34} \left[ \Gamma ^2 (1-a_{21}-a_{43} + a_{21}a_{43}- a_{23}a_{41})+2 \gamma_t  \Gamma  (2-a_{21}-a_{43})+4 \gamma_t ^2 \right] \}
}
\label{nsolutions}
\end{eqnarray}


\section*{References}

\begin{thebibliography}{10}
\bibitem{note1}
With multiple measurements the associated shift will manifest itself as
  statistical variations.
\end{thebibliography}

%

\end{document}